\magnification=\magstep1
\baselineskip=20pt
\centerline{Jets from Collapsing Bubbles}
\bigskip
\centerline{J. I. Katz}
\centerline{Department of Physics and McDonnell Center for the Space
Sciences}
\centerline{Washington University, St. Louis, Mo. 63130}
\centerline{katz@wuphys.wustl.edu}
\bigskip
\centerline{Abstract}
\medskip
When an asymmetric bubble collapses it generally produces a well defined
high velocity jet.  This is remarkable because one might expect such a
collapse to produce a complex or chaotic flow rather than an ordered one.
I present a dimensional argument for the ubiquity of jets from collapsing 
bubbles, and model the aspherical collapse of a bubble with pieces of
Rayleigh's solution for spherical collapse and its cylindrical analogue.
This model explains the ubiquity of jet formation in aspherical collapse,
and predicts the shape and velocity profile of the resulting jet.  These
predictions may be tested in the laboratory or by numerical calculation.  An
application to solid spall is suggested.
\vfill
\noindent
\eject
\centerline{I. Introduction}
\medskip
The aspherical collapse of a bubble or void in a liquid produces a fast 
liquid jet$^{1-3}$.  This phenomenon is remarkably robust.  It occurs for 
axially symmetric collapse of a single bubble near a solid wall or free
surface.  It occurs in at least some of the bubbles produced in turbulent
cavitating flow, but it is apparently unknown whether it occurs in all such
bubbles, or what initial conditions are required.  

Jet production is of great technological importance.  Jets are the means by
which cavitation damages nearby solid surfaces$^{2,4}$.  Fast jets are 
deliberately produced by shaped charges$^5$, and are remarkably insensitive 
to their geometry.  Jets launch droplets from the sea surface$^6$,
producing marine salt aerosols.  Jets are also responsible for the
sensitization of explosives by microscopic bubbles$^{7-8}$.

There are a number of elegant analytic theories of jet production $^{9-11}$.
Numerical calculations of axisymmetric aspherical collapse$^{12-15}$ readily
show jet formation.  However successful these theories and calculations,
they do not explain the robustness of the phenomenon and the applicability
of these somewhat idealized results to the bubbles encountered in practice:
jets form from the collapse of bubbles which cannot be expected to be
symmetric, and despite the best efforts of engineers to prevent them.  A
qualitative model of jet formation might help understand why it is so
ubiquitous.

A simple analytic solution is possible for the collapse of a spheroidal
bubble.  Laplace's equation for the velocity potential $\psi$ separates in
spheroidal coordinates$^{16}$ (either prolate or oblate) $(u,v,\theta)$.
A spheroidal bubble is characterized by $\psi = \psi (u)$.  It is readily 
seen by explicit differentiation that its aspect ratio does not change
during its collapse, so that if it remains spheroidal it will not produce a
jet.  More generally, no bubble which possesses inversion symmetry will
produce a jet, because any jet would stagnate against its mirror image jet
upon convergence.  In fact, inversion symmetry may be broken by the presence
of a nearby wall or free surface, or by the growth of small perturbations,
and initially spherical or ellipsoidal bubbles do produce jets, as is seen
in the numerical calculations.
\bigskip
\centerline{II. Why Jets}
\medskip
A dimensional argument can be made for the ubiquity of jet formation.
Suppose the collapsing bubble is initially approximately spherical, so that
at each point on its surface the two radii of curvature are comparable
to each other and have roughly the same magnitude everywhere on the surface.
Then only one quantity with the dimensions of length (the approximate
initial radius) is defined.  The only other independent dimensional
quantity is a velocity $c \equiv (P/\rho)^{1/2}$, where $\rho$ is the
liquid density and $P$ the pressure at infinity.  If viscosity and surface
tension are neglected and the bubble contains no uncondensable gas there
are also no characteristic dimensionless numbers.  For some geometries
(spheres and spheroids, for example) void collapse will be self-similar,
maintaining the shape of the bubble.  

At a specified elapsed time $t$ a new length scale $ct$ is defined which 
is characteristic of that time, but not of the collapse process as a 
whole.  If the bubble shape is to undergo a qualitative change (such as 
the formation of a jet) its description would require at least one 
additional characteristic length $r^\prime(t)$, typically a radius of 
curvature.  If $r^\prime(t)$ is time-dependent it may be constructed 
from $ct$.  However, it is not possible to define an additional constant 
$r_0^\prime$ which is characteristic of the process as a whole (rather 
than a specific time), because the initial conditions do not contain 
enough information; a limiting, final or time-independent radius of 
curvature would be an example of such a forbidden parameter.

A spherical vacuum bubble satisfies this condition by collapsing to a 
point, rather than reversing its collapse at a finite radius $r_0^\prime$.  
If a collapsing void has an asymmetry or a dimple or pimple on its wall 
and does not preserve its proportions, the asymmetry must either decay or 
sharpen until the flow becomes singular and a cusp forms.  For this reason 
a growing asymmetry will generally lead to a jet which develops singular 
conditions at its tip.  This argument for jet formation also applies to 
bubbles near walls or other bubbles (usually the source of asymmetry), if
all the initial characteristic lengths are comparable to the collapsing 
bubble's initial radius.
\bigskip
\centerline{III. Model}
\medskip
I suggest the following model of jet production:  If the two principal radii
of curvature of a bubble are nearly equal, aspherical collapse is locally
approximated by spherical collapse, as described by a modified
version$^{17}$ of Rayleigh's classic theory$^{18}$, while if the principal
radii of curvature are very different it is locally approximated by
cylindrical collapse.  Collapse of a finite angular range of a cylinder
produces a sheet jet (as in a linear shaped charge$^5$) rather than the
linear jet produced by axially symmetric collapse of an entire
cylinder$^{13}$.  

In this elementary model the difference between spherical (or cylindrical)
collapse and that of an aspherical bubble is that in the aspherical case
different portions of the surface converge to the center at different times.
Instead of meeting an oppositely directed convergent flow from the other
side, and stagnating against it in a central pressure peak (as happens in 
inversion-symmetric collapse), in the asymmetric case the fluid which
converges first forms a fluid jet which then penetrates the unconverged
fluid approaching from the opposite side.  Jets are likely to be produced by
the collapse of any bubble without inversion symmetry.

This model is applicable not only to bubbles and to voids in explosives, but
also to hemispherical shaped charges.  It is not applicable to conical
shaped charges, which are not locally spherical, and whose convergence is
not even locally cylindrical at their apices; a cone defines no quantities
with the dimensions of length, and contains a geometric singularity in its
initial state.

In the frequently encountered case of a bubble near a plane solid boundary
or free surface the collapse is azimuthally symmetric about the surface
normal, and the spherical solution is applicable to the fastest-collapsing
portion of the bubble.  The Rayleigh solution for the velocity field 
surrounding a spherical void, which has 
collapsed from an initial radius $R_0$ to a radius $R$, is
$$v(r) = \left[{2 \over 3}{P \over \rho}\left(R_0^3 - R^3\right)R\right]^{1
/2} r^{-2} \equiv {C_3 \over r^2} \quad (r \ge R), \eqno(1)$$
where here $P$ is the difference between the pressure at infinity and the
pressure in the void.  In the limit $R/R_0 \to 0$, 
$C_3 \to [2 P R_0^3 R / (3 \rho)]^{1/2}$.  The distribution of mass in a
spherical cap of solid angle $\Omega$ with 
respect to specific kinetic energy ${\cal E} \equiv \rho v^2/2$ is then
$${dM \over d{\cal E}} = {\Omega \over 2^{11/4}}{\rho^{7/4} C_3^{3/2} \over
{\cal E}^{7/4}} \quad \left({\cal E} \le {\rho C_3^2 \over 2 R^4}\right).
\eqno(2)$$
Performing the integral $\int {\cal E}(M)\, dM \sim {\cal E}^{1/4}$
demonstrates that the kinetic energy is weakly concentrated in the fluid
with the greatest specific kinetic energy; that is, at the tip of the jet.

During collapse of a cap spherical convergence is assumed (otherwise (1) and
(2) would not be applicable), but after convergence this can no longer be 
the case.  I assume that the fluid then forms a parallel jet, with the 
distribution of speed and kinetic energy given by (1) and (2).  This is not 
required by any conservation law, even for a perfect fluid, but is the 
simplest possible assumption.  It is plausible for a cap of small $\Omega$
because the convergent velocities are nearly parallel and are readily
collimated, and because in a narrow jet the zero pressure boundary
condition along its sides ensures that any longitudinal pressure gradient
and acceleration are small.  

It is necessary to introduce an upper cutoff ${\cal E}_{max}$ on $\cal E$ 
(or, equivalently, a lower cutoff $R_{min} = \left(P/3 {\cal E}_{max}
\right)^{1/3} R_0$ on $R$), because otherwise all the kinetic energy would
appear in an infinitesimal mass of fluid.  This cutoff may be the 
consequence of the onset of compressibility (surface tension and viscosity 
are readily verified to be negligible in the converged flow if they were 
negligible in the original bubble) or a breakdown in the geometric 
assumptions.  The resulting value
$$C_3 = \left({2 \over \rho}\right)^{1/2} \left({P R_0^3 \over 3}\right)^{2/3}
{\cal E}_{max}^{-1/6} \eqno(3)$$
is fortunately only weakly dependent on ${\cal E}_{max}$; the limiting speed
$v_{max} \equiv (2 {\cal E}_{max}/\rho)^{1/2}$.  For water, plausible values
(assuming compressibility is the limiting mechanism) are\break ${\cal E}_{max}
\sim 2 \times 10^{10}$ erg/cm$^3$ and $v_{max} \sim 2 \times 10^5$ cm/sec; 
$R_{min} \sim 0.026 R_0$ if $P = 1$ bar.  The high velocity tip of the jet may
be difficult to observe, because it is eroded by residual gas in the bubble
and by more slowly converging fluid on the opposite side of the bubble.

As the jet propagates it stretches.  If its convergence occurs
instantaneously and at one point then its radius $s$ at a distance $\ell$
from that point at a time $t$ after convergence is
$$s = \left({\Omega \over 2 \pi}\right)^{1/2}{C_3^{3/4} t^{3/4} \over
\ell^{5/4}} \quad (\ell < v_{max}t), \eqno(4)$$
which is obtained by changing variables in (2) from $\cal E$ to $v$ and 
using $\ell = vt$; the jet terminates at $\ell \sim v_{max}t$.  This form is
easier to test against laboratory data than (2) because it is easier to 
measure the shape of a bounding surface than a fluid velocity.

In the case of cylindrical symmetry the solution analogous to (1) for the
velocity field is
$$v(r) = \left[ {P \over \rho}{(R_0^2 - R^2) \over \ln(R_\infty/R)} 
\right]^{1/2} r^{-1} \equiv {C_2 \over r} \quad (r \ge R), \eqno(5)$$
where $R_\infty$ is an upper cutoff (set by the system size) on the range of
the velocity field.  In the limit $R/R_0 \to 0$, $C_2 \to \{PR_0^2/[\rho 
\ln(R_\infty/R)]\}^{1/2}$.  The distribution of mass with respect to $\cal E$ is
$${dM \over d{\cal E}} = {\theta \over 4}{\rho^2 C_2^2 \over {\cal E}^2}
\quad \left({\cal E} \le {\rho C_2^2 \over 2R^2}\right), \eqno(6)$$
where $\theta$ is the arc of the collapsing portion of a cylinder.  The 
integral $\int {\cal E}(M)\, dM \sim \ln {\cal E}$, so that kinetic energy 
is evenly distributed per decade across the specific energy spectrum.  An
upper cutoff ${\cal E}_{max}$ and a lower cutoff $R_{min}$ are again required
as $R \to 0$.  The thickness $h$ of a collapsed sheet is found, in analogy 
to (4),
$$h = \theta{C_2^2 t^2 \over \ell^3} \quad (\ell < v_{max}t). \eqno(7)$$

Collapsing bubbles whose rate of convergence is intermediate between
cylindrical and spherical in their region of fastest collapse may perhaps be
parametrized by solutions of non-integer dimension $n$.  The velocity field is
$$v(r) = \left[{2(n - 2) \over n}{P \over \rho}(R_0^n - R^n)R^{n-2}
\right]^{1/2} r^{1-n} \equiv {C_n \over r^{n-1}} \quad (r \ge R). \eqno(8)$$
The resulting mass distribution in the limit $R/R_0 \to 0$ is
$${dM \over d{\cal E}} \propto \left({\rho \over {\cal E}} \right)^{(3n - 2)
/ (2n - 2)} C_n^{n/(n - 1)} \quad \left({\cal E} \le {\rho C_n^2 \over 2
R^{2(n-1)}}\right), \eqno(9)$$
where the constant of proportionality includes the contributing fraction of
the $n$-sphere.  The shape of the jet's cross-section depends on the details of 
convergence, but with the previous assumptions its cross-sectional area $A$ is
$$A \propto {t^{n/(n - 1)} \over \ell^{(2n - 1)/(n - 1)}} \quad (\ell <
v_{max}t). \eqno(10)$$
This may be fitted to empirical data or to numerical calculation to
determine an effective dimension $n$ of the convergent flow.
\bigskip
\centerline{IV. Discussion}
\medskip
The models of jets discussed in this paper can be tested by comparison to
computed jets and to experiment.  The most general form is (10), which 
introduces the non-integer dimensionality parameter $n$, but which reduces 
to the spherical results (1)--(4) for $n=3$ and to the cylindrical results
(5)--(7) for $n=2$.  

A related problem is the production of microscopic particulate spall upon shock 
reflection from a solid surface, at tensile loads insufficient to disrupt the 
bulk.  This is related to fluid jet formation, because both processes involve 
concentration of energy.  Solid spall is a more complicated phenomenon
because it involves materials with finite strength, a variety of heterogeneities
in the bulk and at the surface, and (usually) anisotropy.
It is unclear whether spall is produced by elastic stress concentration at 
corners (surface scratches, cracks, grain boundaries, {\it etc.}), followed by 
brittle fracture, or by plastic flow convergence and jetting at surface 
scratches and cracks.  The latter process would resemble jet formation upon
the collapse of a bubble, with the curved solid surface taking the place of
the bubble surface.  The plastic flow and brittle fracture hypotheses may be
distinguished by microscopic examination of the surfaces of spall fragments.
It might also be informative to do experiments on spall from shocked liquid 
surfaces and amorphous substances, which may be prepared without surface 
imperfections or heterogeneities in the bulk.

I thank K. Case and F. J. Dyson for discussions and the Office of Naval
Research, DARPA and NSF AST 94-16904 for support.
\vfil
\eject
\centerline{References}
\medskip
\item{1.} M. Van Dyke, {\it Album of Fluid Motion} (Parabolic Press,
Stanford, Cal., 1982).
\item{2.} F. R. Young, {\it Cavitation} (McGraw-Hill, London, 1989).
\item{3.} R. H. Cole, {\it Underwater Explosions} (Princeton U. Press, 
Princeton, 1948).
\item{4.} J. R. Blake and D. C. Gibson, ``Cavitation bubbles near
boundaries'' {\it Ann. Rev. Fluid Mech.} {\bf 19}, 99--123 (1987).
\item{5.} W. P. Walters and J. A. Zukas, {\it Fundamentals of Shaped
Charges} (Wiley, New York, 1989).
\item{6.} D. C. Blanchard and A. H. Woodcock, ``The production, concentration
and vertical distribution of the sea-salt aerosol'' {\it Ann. N. Y. Acad.
Sci.} {\bf 338}, 330--347 (1980).
\item{7.} C. Matzner, {\it Physics Today} {\bf 46}(4), 90 (1993).
\item{8.} M. M. Chaudhri, {\it Physics Today} {\bf 45}(7), 15 (1992); {\bf
46}(4), 90 (1993).
\item{9.} M. S. Longuet-Higgins, ``Bubbles, breaking waves and hyperbolic
jets at a free surface'' {\it J. Fluid Mech.} {\bf 127}, 103--121 (1983).
\item{10.} M. S. Longuet-Higgins, ``Inertial shocks in surface waves and
collapsing bubbles'' {\it Proc. IUTAM Symp. on Bubble Dynamics and Interface
Phenomena}, Birmingham, England (1993).
\item{11.} M. S. Longuet-Higgins and H. Oguz, ``Critical microjets in
collapsing cavities'' {\it J. Fluid Mech.} {\bf 290}, 183--201 (1995).
\item{12.} M. S. Plesset and R. B. Chapman, ``Collapse of an initially
spherical vapour cavity in the neighbourhood of a solid boundary'' {\it J. 
Fluid Mech.} {\bf 47}, 283--290 (1971).
\item{13.} O. V. Voinov and V. V. Voinov, ``On the process of collapse of a
cavitation bubble near a wall and the formation of a cumulative jet'' {\it
Sov. Phys. Dokl.} {\bf 21}, 133--135 (1976).
\item{14.} J. R. Blake, B. B. Taib and G. Doherty, ``Transient cavities near
boundaries. Part 1. Rigid boundary'' {\it J. Fluid Mech.} {\bf 170}, 479--497
(1987).
\item{15.} J. R. Blake, B. B. Taib and G. Doherty, ``Transient cavities near
boundaries. Part 2. Free surface'' {\it J. Fluid Mech.} {\bf 181}, 197--212
(1987).
\item{16.} F. B. Hildebrand, {\it Advanced Calculus for Engineers} 
(Prentice-Hall, New York, 1949).
\item{17.} W. V. Pinczewski, ``The formation and growth of bubbles at a
submerged orifice'' {\it Chem. Eng. Sci.} {\bf 36}, 405--411 (1981).
\item{18.} H. Lamb, {\it Hydrodynamics} 6th ed. (Dover, New York, 1945).
\vfil
\eject
\end